\documentstyle[12pt]{article}
\input epsf.tex

\begin{document}

\baselineskip=15.5pt
\pagestyle{plain}
\setcounter{page}{1}

\renewcommand{\thefootnote}{\fnsymbol{footnote}}

\begin{titlepage}

\begin{flushright}
PUPT-1831\\
hep-th/9901018
\end{flushright}
\vfil

\begin{center}
{\huge From Threebranes}\\
\vspace{3 mm}
{\huge to Large N Gauge Theories
\footnote{Based on the talk at Orbis Scientiae '98, Ft. Lauderdale,
Florida, December 17-21, 1998.}}
\end{center}

\vfil

\begin{center}
{\large Igor R.\ Klebanov}\\
\vspace{3mm}
Joseph Henry Laboratories\\
Princeton University\\
Princeton, New Jersey 08544, USA\\
\vspace{3mm}
\end{center}

\vfil

\begin{center}
{\large Abstract}
\end{center}

\noindent
This is a brief introductory review of the AdS/CFT correspondence
and of the ideas that led to its formulation. Emphasis is placed
on dualities between conformal large $N$ gauge theories in
4 dimensions and string backgrounds of the form $AdS_5\times X_5$.
Attempts to generalize this correspondence to
asymptotically free theories are also included.
\vfil
\begin{flushleft}
January 1999
\end{flushleft}
\end{titlepage}
\newpage

\renewcommand{\thefootnote}{\arabic{footnote}}
\setcounter{footnote}{0}


\newcommand{\grad}{\nabla}
\newcommand{\tr}{\mathop{\rm tr}}
\newcommand{\half}{{1\over 2}}
\newcommand{\third}{{1\over 3}}
\newcommand{\be}{\begin{equation}}
\newcommand{\ee}{\end{equation}}
\newcommand{\bea}{\begin{eqnarray}}
\newcommand{\eea}{\end{eqnarray}}

\newcommand{\dint}[2]{\int\limits_{#1}^{#2}}
\newcommand{\D}{\displaystyle}
\newcommand{\PDT}[1]{\frac{\partial #1}{\partial t}}
\newcommand{\PD}{\partial}
\newcommand{\tw}{\tilde{w}}
\newcommand{\tg}{\tilde{g}}
\newcommand{\newcaption}[1]{\centerline{\parbox{6in}{\caption{#1}}}}
\def\href#1#2{#2}  

\def \ci {\cite}
\def \foot {\footnote}
\def \bi{\bibitem}
\newcommand{\rf}[1]{(\ref{#1})}
\def \del{\partial}
\def \m {\mu}
\def \n {\nu} 
\def \g {\gamma}
\def \G {\Gamma}
\def \a {\alpha}
\def \ov {\over}
\def \la {\label}
\def \ep {\epsilon}
\def \d {\delta}
\def \k {\kappa}
\def \p {\phi}
\def \ha {\textstyle{1\ov 2}}

\def\np {  {\em Nucl. Phys.} }
\def \pl { {\em Phys. Lett.} }
\def \mpl { Mod. Phys. Lett. }
\def \prl { Phys. Rev. Lett. }
\def \pr  { {\em Phys. Rev.} }
\def \cqg { Class. Quantum Grav.}
\def \jmp { Journ. Math. Phys. }
\def\ap { Ann. Phys. }
\def \ijmp { Int. J. Mod. Phys. }

%
\def\TL{\hfil$\displaystyle{##}$}
\def\TR{$\displaystyle{{}##}$\hfil}
\def\TC{\hfil$\displaystyle{##}$\hfil}
\def\TT{\hbox{##}}
\def\seqalign#1#2{\vcenter{\openup1\jot
  \halign{\strut #1\cr #2 \cr}}}

\def\comment#1{}
\def\fixit#1{}

\def\tf#1#2{{\textstyle{#1 \over #2}}}
\def\df#1#2{{\displaystyle{#1 \over #2}}}

\def\mop#1{\mathop{\rm #1}\nolimits}

\def\ad{\mop{ad}}
\def\coth{\mop{coth}}
\def\csch{\mop{csch}}
\def\sech{\mop{sech}}
\def\Vol{\mop{Vol}}
\def\vol{\mop{vol}}
\def\diag{\mop{diag}}
\def\tr{\mop{tr}}
\def\Disc{\mop{Disc}}
\def\sgn{\mop{sgn}}

\def\SU{{\rm SU}}
\def\USp{{\rm USp}}            

\def\lsim{\mathrel{\mathstrut\smash{\ooalign{\raise2.5pt\hbox{$<$}\cr\lower2.5pt\hbox{$\sim$}}}}}
\def\gsim{\mathrel{\mathstrut\smash{\ooalign{\raise2.5pt\hbox{$>$}\cr\lower2.5pt\hbox{$\sim$}}}}}

\def\slashed#1{\ooalign{\hfil\hfil/\hfil\cr $#1$}}

\def\sqr#1#2{{\vcenter{\vbox{\hrule height.#2pt
         \hbox{\vrule width.#2pt height#1pt \kern#1pt
            \vrule width.#2pt}
         \hrule height.#2pt}}}}
\def\square{\mathop{\mathchoice\sqr56\sqr56\sqr{3.75}4\sqr34\,}\nolimits}

\def\idget{$\sqr55$\hskip-0.5pt}
\def\endrow{\hskip0.5pt\cr\noalign{\vskip-1.5pt}}
\def\endyoung{\hskip0.5pt\cr}

\def\href#1#2{#2}  


%
\def\lbldef#1#2{\expandafter\gdef\csname #1\endcsname {#2}}
\def\eqn#1#2{\lbldef{#1}{(\ref{#1})}%
\begin{equation} #2 \label{#1} \end{equation}}
\def\eqalign#1{\vcenter{\openup1\jot
    \halign{\strut\span\TL & \span\TR\cr #1 \cr
   }}}
\def\eno#1{(\ref{#1})}

\def\rmax{{r_{\rm max}}}
\def\gone#1{}

\section{Introduction}

It is well-known that string theory originated from attempts to understand 
the strong interactions \cite{NNS}. However, after the emergence of QCD
as the theory of hadrons, the dominant theme of string research
shifted to the Planck scale domain of quantum gravity \cite{Scherk}. 
Although in hadron physics one routinely hears about flux tubes
and the string tension, the majority of particle theorists gave up hope
that string theory might lead to an exact description 
of the strong interactions.
Now, however, for the first time we can say with confidence
that at least some strongly coupled gauge theories have a dual description
in terms of strings. Let me emphasize that one is not talking here about
effective strings that give an approximate qualitative description, but
rather about an exact duality. At weak coupling a convenient
description of the theory involves conventional perturbative methods;
at strong coupling, where such methods are intractable, the dual string
description simplifies and gives exact information about the theory.
The best established examples of this duality are conformal gauge theories
where the so-called AdS/CFT correspondence \cite{jthroat,US,EW}
has allowed for many calculations
at strong coupling to be performed with ease. In these notes I describe,
from my own personal perspective, some of the ideas that led to the formulation
of the AdS/CFT correspondence. I will also speculate on the future
directions.  For the sake of brevity I will mainly discuss the
AdS$_5$/CFT$_4$ case which is most directly related to 4-dimensional
gauge theories.

It has long been believed that the best hope for a string description
of non-Abelian gauge theories lies in the 't~Hooft large $N$ limit.
A quarter of a century ago 't~Hooft proposed to generalize the $SU(3)$
gauge group of QCD to $SU(N)$, and to take the large $N$ limit while keeping
$g_{{\rm YM}}^2 N$ fixed \cite{GT}. In this limit each Feynman graph carries
a topological factor $N^\chi$, where $\chi$ is the Euler
characteristic of the graph.
Thus, the sum over graphs of a given topology can perhaps be thought of
as a sum over world sheets of a hypothetical ``QCD string.'' 
Since the spheres (string tree diagrams)
are weighted by $N^2$, the tori (string one-loop diagrams) -- 
by $N^0$, etc., we find that the closed string coupling constant is of order
$N^{-1}$. Thus, the advantage of taking $N$ to be large is that we find
a weakly coupled string theory. It is not clear, however, how to describe
this string theory in elementary terms (by a 2-dimensional world sheet action, 
for example). This is clearly an important problem: the 
free closed string spectrum is just the large $N$ spectrum of glueballs.
If the quarks are included, then we also find open strings describing the 
mesons. Thus, if methods are developed for calculating these spectra,
and it is found that they are discrete, then
this provides an elegant
explanation of confinement. Furthermore, the $1/N$ corrections 
correspond to perturbative string corrections.

Many years of effort, and many good ideas, were invested into the search
for an exact gauge field/string duality \cite{book}. 
One class of ideas, exploiting
the similarity of the large $N$ loop equation with the string Schroedinger
equation, eventually led to the following fascinating 
speculation \cite{Sasha}:
one should not look for the QCD string in four dimensions, but rather
in five, with the fifth dimension akin to the Liouville dimension of
non-critical string theory \cite{bos}. This leads to a picture
where the QCD string is described by a two-dimensional world sheet
sigma model with a curved 5-dimensional target space. 
At that stage it was not clear,
however, precisely what target spaces are relevant to gauge theories.
Luckily, we now do have answers to this question for a variety of
conformal large $N$ gauge models. The route that leads to this answer,
and confirms the idea of the fifth dimension,
involves an unexpected detour via black holes and Dirichlet branes.
We turn to these subjects next.

\section{D-branes vs. Black Holes and $p$-branes}

A few years ago it became clear that, 
in addition to strings, superstring theory
contains soliton-like ``membranes'' of various internal dimensionalities
called Dirichlet branes (or D-branes) \cite{Dnotes}.
A Dirichlet $p$-brane (or D$p$-brane) is a $p+1$ dimensional hyperplane
in $9+1$ dimensional space-time where strings are allowed to end,
even in theories where all strings are closed in the bulk of space-time.
In some ways a D-brane is like a topological defect: when a closed string
touches it, it can open open up and turn into an open string whose
ends are free to move along the D-brane. For the end-points of such a string
the $p+1$ longitudinal coordinates satisfy the conventional free (Neumann)
boundary conditions, while the $9-p$ coordinates transverse to
the D$p$-brane have the fixed (Dirichlet) boundary conditions; hence
the origin of the term ``Dirichlet brane.'' 
In a seminal paper \cite{brane} Polchinski
showed that the D$p$-brane is a BPS saturated object which preserves
$1/2$ of the bulk supersymmetries and carries an elementary unit
of charge with respect to the $p+1$ form gauge potential from the
Ramond-Ramond sector of type II superstring. The existence of BPS
objects carrying such charges is required by non-perturbative string
dualities \cite{HullT}. A striking feature of the D-brane formalism 
is that it provides a
concrete (and very simple) embedding of such objects into perturbative
string theory.

Another fascinating feature of the D-branes is that they naturally realize
gauge theories on their world volume. The massless spectrum of open strings
living on a D$p$-brane is that of a maximally supersymmetric $U(1)$
gauge theory in $p+1$ dimensions. The $9-p$ massless
scalar fields present in this supermultiplet are the expected Goldstone 
modes associated with the transverse oscillations of the D$p$-brane,
while the photons and fermions
may be thought of as providing the unique supersymmetric
completion.
If we consider $N$ parallel D-branes,
then there are $N^2$ different species of open strings because they can
begin and end on any of the D-branes. 
$N^2$ is the dimension of the adjoint representation of $U(N)$,
and indeed we find the maximally supersymmetric $U(N)$ 
gauge theory in this setting \cite{Witten}.
The relative separations of the D$p$-branes in the $9-p$ transverse
dimensions are determined by the expectation values of the scalar fields.
We will be primarily interested in the case where all scalar expectation
values vanish, so that the $N$ D$p$-branes are stacked on top of each other.
If $N$ is large, then this stack is a heavy object embedded into a theory
of closed string which contains gravity. Naturally, this macroscopic
object will curve space: it may be described by some classical metric
and other background fields, such as the Ramond-Ramond 
$p+1$ form potential.
Thus, we have two very different descriptions of the stack of D$p$-branes:
one in terms of the $U(N)$ supersymmetric gauge theory on its world volume,
and the other in terms of the classical
Ramond-Ramond charged $p$-brane background of the type II
closed superstring theory. It is the relation between these two descriptions
that is at the heart of the recent progress in understanding the
connections between gauge fields and strings.\footnote{
There are other similar relations between large $N$ SYM theories
and gravity stemming from the BFSS matrix theory conjecture
\cite{bfss}.} Of course, more work is needed to make this relation precise.

\subsection{Counting the entropy}

The first success in building this kind of correspondence between
black hole metrics and D-branes was achieved by Strominger and Vafa
\cite{SV}. They considered 5-dimensional
supergravity obtained by compactifying 10-dimensional type IIB theory
on a 5-dimensional compact manifold (for example, the 5-torus), 
and constructed a class of black holes carrying
2 separate $U(1)$ charges. These solutions may be viewed as generalizations
of the well-known 4-dimensional charged (Reissner-Nordstrom) black hole.
For the Reissner-Nordstrom black hole the mass
is bounded from below by a quantity proportional to the charge.
In general, when the mass saturates the lower (BPS)
bound for a given choice
of charges, then the black hole is called extremal.
The extremal Strominger-Vafa black hole
preserves $1/8$ of the supersymmetries present in vacuum.
Also, the black hole is constructed
in such a way that, just as for the Reissner-Nordstrom solution, 
the area of the
horizon is non-vanishing at extremality \ci{SV}. 
In general, an important quantity characterizing
black holes is the Bekenstein-Hawking entropy which is proportional
to the horizon area:
\be S_{BH} = {A_h\over 4G}
\ ,
\ee
where $G$ is the Newton constant.
Strominger and Vafa calculated the Bekenstein-Hawking entropy of their
extremal solution as a function of the charges and succeeded in reproducing
this result with D-brane methods. To build a D-brane system carrying the 
same set of charges as the black hole, they had to consider intersecting
D-branes wrapped over the compact 5-dimensional manifold. 
For example, one may consider D3-branes intersecting over a line or
D1-branes embedded inside D5-branes. The $1+1$ dimensional gauge theory
describing such an intersection is quite complicated, but the degeneracy
of the supersymmetric BPS states can nevertheless be calculated in the
D-brane description valid at weak coupling. 
For reasons that will become clear shortly,
the description in terms of black hole metrics is valid only at very
strong coupling. Luckily, due to the supersymmetry,
the number of states does not change as the coupling is increased.
This ability to extrapolate the D-brane counting to strong coupling
makes a comparison with the Bekenstein-Hawking entropy possible,
and exact agreement is found in the limit of large charges \cite{SV}.
In this sense the collection of D-branes provides a ``microscopic''
explanation of the black hole entropy.

This correspondence was quickly generalized to black 
holes slightly excited above the extremality \cite{cm,HSS}. 
Further, the Hawking radiation rates and the absorption cross-sections
were calculated and successfully reproduced by D-brane models
\cite{cm,wad,ms}.
Since then this system has been
receiving a great deal of attention.
However, some detailed comparisons are hampered by the complexities of the
dynamics of intersecting D-branes: to date there is no 
first principles approach to the lagrangian of the $1+1$ dimensional 
conformal field theory on the intersection. 

For this and other reasons it has turned out 
very fruitful to study a similar correspondence
for simpler systems which involve parallel D-branes 
only \cite{gkp,ENT,kleb,gukt,gkThree}.
Our primary motivation is that, as explained above, parallel
D$p$-branes realize $p+1$ dimensional $U(N)$ SYM theories, and we may
learn something new about them from comparisons with Ramond-Ramond
charged black $p$-brane
classical solutions. These solutions in type II supergravity have been
known since the early 90's \cite{hs,DL}.
The metric and dilaton backgrounds may be expressed in the following
simple and universal form:
\be
\label{metric}
   ds^2 =
H^{-1/2}(r)
    \left[ - f(r) dt^2 + \sum_{i=1}^p (d x^i)^2 \right] +
H^{1/2}(r)
    \left[f^{-1} (r) dr^2 + r^2 d\Omega_{8-p}^2 \right] \ ,
\ee
$$ e^\Phi = H^{(3-p)/4}(r)\ ,
$$
where
$$   H(r)  = 1 + {L^{7-p} \over r^{7-p}} \ , \qquad \  \  
f(r) = 1- {r_0^{7-p}\over
r^{7-p}}
\ ,
$$
and $d\Omega_{8-p}^2$ is the metric of a unit $8-p$ dimensional sphere.
The horizon is located at $r=r_0$ and the extremality
is achieved in the limit $r_0 \rightarrow 0$.
A solution with $r_0\ll L$ is called near-extremal.
In contrast to the 
situation encountered for the Strominger-Vafa black hole,
the Bekenstein-Hawking entropy vanishes in the extremal limit.
Just like the stacks of parallel D-branes, the extremal solutions are 
BPS saturated: they preserve 16 of the 32 supersymmetries present
in the type II theory. For $r_0>0$ the $p$-brane carries some excess
energy $E$ above its extremal value, and the
Bekenstein-Hawking entropy is also non-vanishing. 
The Hawking temperature is then defined by
$ T^{-1} = \partial S_{BH}/\partial E$.

The correspondence between the entropies of the
$p$-brane solutions (\ref{metric}) and those
of the $p+1$ dimensional SYM theories was first considered in
\cite{gkp,ENT}. Among these solutions 
$p=3$ has a special status: in the extremal limit $r_0 \rightarrow 0$
the 3-brane solution is perfectly non-singular \cite{gt}.
This is evidenced by the fact that the dilaton $\Phi$ is constant for
$p=3$, but blows up at $r=0$ for all other extremal solutions.
In \cite{gkp} the Bekenstein-Hawking entropy of a near-extremal
3-brane of Hawking temperature $T$ was compared with the entropy
of the ${\cal N}=4$ supersymmetric $U(N)$ gauge theory (which lives
on $N$ coincident D3-branes) heated up to the same temperature. 
The results turned out to be quite interesting.
The Bekenstein-Hawking entropy expressed in terms of the
Hawking temperature $T$ and the number $N$ of elementary units of charge
was found to be
\begin{equation}
\label{bhe}
S_{BH}= {\pi^2\over 2} N^2 V_3 T^3
\ ,
\end{equation}
where $V_3$ is the spatial volume of the 3-brane.
This was compared with the entropy of a free $U(N)$ ${\cal N}=4$
supermultiplet, which consists of the gauge field, $6 N^2$ massless
scalars and $4 N^2$ Weyl fermions. This entropy was calculated using
the standard statistical mechanics of a massless gas (the black body 
problem), and the answer turned out to be
\be S_0= {2 \pi^2\over 3} N^2 V_3 T^3
\ .
\ee
It is remarkable that the 3-brane geometry
captures the $T^3$ scaling characteristic of a conformal
field theory (in a CFT this scaling is guaranteed by the extensivity of
the entropy and the absence of dimensionful parameters).\footnote{
Other examples of the ``conformal'' behavior of the
Bekenstein-Hawking entropy include the 11-dimensional 5-brane and
membrane solutions \cite{ENT}. For the 5-brane,
$S_{BH}\sim N^3 T^5 V_5$, while for the membrane
$S_{BH}\sim N^{3/2} T^2 V_2$. The microscopic description of
the 5-brane solution is in terms of a large number $N$ of coincident singly
charged 5-branes of M-theory, whose chiral world volume theory has
$(0,2)$ supersymmetry. Similarly, the membrane solution 
describes the large $N$ behavior of the CFT on $N$ coincident
elementary membranes. The entropy formulae suggest that these theories
have $O(N^3)$ and $O(N^{3/2})$ massless degrees of freedom respectively.
These predictions of supergravity \cite{ENT}
are non-trivial and still mysterious. Since the geometry
of the 5-brane throat is $AdS_7\times S^4$, and that of the 
membrane throat is $AdS_4\times S^7$, these systems lead to other
interesting examples of the AdS/CFT correspondence.}
Also, the $N^2$ scaling indicates the presence of $O(N^2)$
unconfined degrees of freedom, which is exactly what we expect in
the ${\cal N}=4$ supersymmetric $U(N)$ gauge theory.
On the other hand, the relative factor of $3/4$ between $S_{BH}$ and
$S_0$ at first appeared mysterious and was interpreted by many as
a subtle failure of the D3-brane approach to black 3-branes.
As we will see shortly, however, the relative factor of $3/4$
is not a contradiction but rather a prediction about strongly
coupled ${\cal N}=4$ SYM theory at finite temperature. 

\subsection{From absorption cross-sections to two-point correlators}

Almost a year after the entropy comparisons
\cite{gkp,ENT} I came back to the 3-branes (and also to the
11-dimensional membranes and 5-branes) and tried to interpret
absorption cross-sections for massless particles
in terms of the world volume theories \cite{kleb}. 
This was a natural step beyond
the comparison of entropies, and for the Strominger-Vafa black holes
the D-brane approach to absorption was initiated earlier in \cite{cm,wad}.
For the system of $N$ coincident D3-branes it was interesting
to inquire to what extent the supergravity and the weakly coupled
D-brane calculations agreed. For example, they might scale differently
with $N$ or with the incident energy. Even if the scaling exponents 
agreed, the overall normalizations
could differ by a subtle numerical factor similar
to the $3/4$ found for the 3-brane entropy.
Surprisingly, the low-energy absorption cross-sections
turned out to agree exactly!

To calculate the absorption cross-sections in the D-brane formalism one
needs the low-energy world volume action for coincident D-branes coupled to
the massless bulk fields. Luckily, these couplings may be deduced
from the D-brane Born-Infeld action. For example, the coupling
of 3-branes to the dilaton $\Phi$, the Ramond-Ramond scalar $C$,
and the graviton $h_{\alpha\beta}$ is given by \cite{kleb,gukt}
\be \label{sint}
   S_{\rm int} = {\sqrt \pi\over\kappa}
\int d^4 x \, \bigg[ \tr \left(
 \tf{1}{4}   {\Phi} F_{\alpha\beta}^2  -
  \tf{1}{4}   {C} F_{\alpha\beta} \tilde {F}^{\alpha\beta} \right)
+ \tf{1}{2} h^{\alpha\beta} T_{\alpha\beta} \bigg] \ ,
 \ee
 where $T_{\alpha\beta}$ is
the stress-energy tensor of the ${\cal N}=4$ SYM theory.
Consider, for instance, absorption of a dilaton incident on the 3-brane
at right angles with a low energy
$\omega$. Since the dilaton couples to $\tr F_{\alpha\beta}^2$
it can be converted into a pair of back-to-back gluons on the world volume.
The leading order calculation of the cross-section
for weak coupling gives \cite{kleb}
\be\label{absorb}
   \sigma = {\kappa^2  \omega^3 N^2\over 32 \pi} \ ,
 \ee
where $\kappa=\sqrt{8\pi G}$ is the 10-dimensional gravitational 
constant (note that the factor $N^2$ comes from the degeneracy of
the final states which is the number of different gluon species).
This result was compared with the absorption cross-section by
the extremal 3-brane geometry,
\be
\label{geom}
ds^2 = \left (1+{L^4\over r^4}\right )^{-1/2}
\left (- dt^2 +dx_1^2+ dx_2^2+ dx_3^2\right )
+ \left (1+{L^4\over r^4}\right )^{1/2}
\left ( dr^2 + r^2 d\Omega_5^2 \right )\ .
\ee
This geometry may be viewed as a semi-infinite throat which for
$r \gg L$ opens up into flat $9+1$ dimensional space.
Waves incident from the $r \gg L$ region partly reflect back and
partly penetrate into the the throat region $r \ll L$.
The relevant s-wave radial equation turns out to be \cite{kleb}
\be
\label{Coulthree}
\left [{d^2\over d \rho^2} - {15\over 4 \rho^2}
+1 + {(\omega L)^4\over \rho^4} \right ] \psi(\rho) =0\ ,
\ee
where $\rho = \omega r$. For a low energy $\omega \ll 1/L$ we find
a high barrier separating the two asymptotic regions.
The low-energy behavior of the tunneling probability may be calculated
by the so-called matching method, and the resulting absorption
cross-section is \cite{kleb}
\be
\label{three}
\sigma_{SUGRA}= {\pi^4\over 8}\omega^3 L^8 \ .
\ee
In order to compare (\ref{absorb}) and (\ref{three}) we need a relation
between the radius of the throat, $L$, and the number of D3-branes, $N$.
Such a relation follows from equating the ADM tension of the extremal
3-brane solution to $N$ times the tension of a single D3-brane, and
one finds \cite{gkp}
\be\label{throatrel}
L^4 = {\kappa\over 2\pi^{5/2}} N \ .
\ee
Substituting this into (\ref{three}), we find that the supergravity
absorption cross-section agrees exactly with the D-brane one,
without any relative factor like $3/4$.

This result was a major surprise to me, and I started searching for
its explanation. The most important question is: what is the range of
validity of the two calculations? Since $\kappa\sim g_{st}
\alpha'^2$, (\ref{throatrel}) gives
$ L^4 \sim N g_{st} \alpha'^2$. Supergravity can only be trusted if
the length scale of the 3-brane solution is much larger than the 
string scale $\sqrt{\alpha'}$,
i.e. for $N g_{st} \gg 1$.\footnote{A similar
conclusion applies to the Strominger-Vafa black hole \cite{SV}.}
Of course, the incident energy also has
to be small compared to $1/\sqrt{\alpha'}$. 
Thus, the supergravity calculation should
be valid in the ``double-scaling limit'' \cite{kleb}
\begin{equation}
\label{dsl}
{L^4\over \alpha'^2} \sim  g_{st} N \rightarrow \infty\ ,
\qquad\qquad \omega^2 \alpha' \rightarrow 0\ .
\end{equation}
If the description of the black 3-brane by a stack of
many coincident D3-branes is correct, and we presume that it is,
then it {\it must} agree with the supergravity results in this limit.
Since $g_{st}\sim g_{\rm YM}^2$, this corresponds to the limit of
{\it infinite} `t Hooft coupling in the  
${\cal N}=4$ $U(N)$ SYM theory. Since we also want to send $g_{st}
\rightarrow 0$ in order to suppress the string loop corrections,
we necessarily have to take the large $N$ limit.
To summarize, the supergravity calculations are expected to give exact
information about the ${\cal N}=4$ SYM theory in the limit of large $N$
and large `t Hooft coupling \cite{kleb}.

Coming back to the entropy problem, we now see that the 
Bekenstein-Hawking entropy calculation
applies to the $g_{\rm YM}^2 N\rightarrow\infty$ limit of the theory,
while the free field calculation applies to the $g_{\rm YM}^2 N\rightarrow
0$ limit. Thus, the relative factor of $3/4$ is not a discrepancy:
it relates two different limits of the theory. 
Indeed, on general grounds we expect that in the `t Hooft large $N$
limit the entropy is given by \cite{GKT}
\be
S= {2 \pi^2\over 3} N^2 f(g_{\rm YM}^2 N) V_3 T^3
\ .\ee
The function $f$ is certainly not constant: for example, a recent
two-loop calculation \cite{foto} shows that its perturbative expansion is
\be 
f(g_{\rm YM}^2 N) = 1 - {3\over 2\pi^2} g_{\rm YM}^2 N + \ldots
\ee
Thus, the Bekenstein-Hawking entropy in supergravity, (\ref{bhe}),
is translated into the prediction that $f(g_{{\rm YM}}^2 N \rightarrow \infty)
=3/4$. In fact, a recent string theory calculation of the leading strong
coupling correction gives \cite{GKT}
\be
 f(g_{{\rm YM}}^2 N) = {3\over 4} +
{45\over 32} \zeta(3) (2 g_{{\rm YM}}^2 N)^{-3/2}  + \ldots
\ .
\ee
This is consistent with $f(g_{{\rm YM}}^2 N)$ being a monotonic function
which interpolates between 1 at $g_{{\rm YM}}^2 N=0$ and
$3/4$ at $g_{{\rm YM}}^2 N=\infty$.

Although we have sharpened the region of applicability of the
supergravity calculation (\ref{three}), we have not yet explained why
it agrees with the leading order perturbative result (\ref{absorb})
on the D3-brane world volume. 
After including the higher-order SYM corrections, the general structure
of the absorption cross-section in the large $N$ limit
is expected to be \cite{gkThree}
\be\label{newabsorb}
   \sigma = {\kappa^2  \omega^3 N^2\over 32 \pi} a(g_{\rm YM}^2 N)\ ,
 \ee
where
$$ a(g_{\rm YM}^2 N) = 1 + b_1 g_{\rm YM}^2 N + 
b_2 (g_{\rm YM}^2 N)^2 + \ldots
$$
For agreement with supergravity, the strong
`t Hooft coupling limit of $a(g_{\rm YM}^2 N)$ should be equal to 1 
\cite{gkThree}. In fact, a stronger result is true: all perturbative
corrections vanish and $a=1$ independent of the coupling.
This was first shown explicitly in \cite{gkThree} for the graviton
absorption. The absorption cross-section is related to the
imaginary part of the two-point function
$\langle T_{\alpha\beta} (p) T_{\gamma\delta} (-p) \rangle$
in the SYM theory. In turn,
this is determined by a conformal ``central charge''
which satisfies a non-renormalization theorem: it is completely
independent of the `t Hooft coupling.

In general, the two-point function of a gauge invariant operator in
the strongly coupled SYM theory may be read off from the
absorption cross-section for the supergravity field which
couples to this operator in the world volume action \cite{gkThree}.
Some examples of this field operator correspondence may
be read off from (\ref{sint}). Thus, we learn that the dilaton
absorption cross-section measures the imaginary part of
$\langle \tr F_{\alpha\beta}^2 (p) \tr F_{\gamma\delta}^2 (-p) \rangle$,
the Ramond-Ramond scalar absorption cross-section measures the 
imaginary part of
$\langle \tr F_{\alpha\beta} \tilde {F}^{\alpha\beta}(p) 
\tr F_{\gamma\delta} \tilde {F}^{\gamma\delta} (-p) \rangle$,
etc. The agreement of these two-point functions
with the weak-coupling calculations performed in
\cite{kleb,gukt} is explained by supersymmetric non-renormalization
theorems. Thus, the proposition that the $g_{\rm YM}^2 N\rightarrow
\infty$ limit of the large $N$ ${\cal N}=4$ SYM theory can be extracted from 
the 3-brane of type IIB supergravity has passed its first consistency checks.

\section{The AdS/CFT Correspondence}

The circle of ideas reviewed in the previous section received
a seminal development by Maldacena \cite{jthroat} who also connected
it for the first time with the QCD string idea. Maldacena made
a simple and powerful observation that the ``universal'' region of
the 3-brane geometry, which should be directly identified with the
${\cal N}=4$ SYM theory, is the throat, i.e. the region $r\ll L$.\footnote{
Related ideas were also pursued in \cite{HYU}.} The
limiting form of the metric (\ref{geom}) is
\be \label{adsmetric}
ds^2 = {L^2 \over z^2} \left( -dt^2 + d\vec{x}^2 + dz^2 \right) +
    L^2 d\Omega_5^2 \ ,
\ee
where $z={L^2\over r}\gg L$. This metric describes the space
$AdS_5\times S^5$ with equal radii of curvature $L$.
One also finds that the self-dual 5-form Ramond-Ramond field strength
has constant flux through this space (the field strength term in the
Einstein equation effectively gives a positive cosmological constant on $S^5$
and a negative one on $AdS_5$).
Thus, Maldacena conjectured that type IIB string theory on $AdS_5\times S^5$
should be somehow dual to the large $N$ ${\cal N}=4$ SYM theory.

Maldacena's argument was based on the fact that the low-energy
($\alpha'\rightarrow 0$) limit may be taken directly in the 3-brane
geometry and is equivalent to the throat ($r \rightarrow 0$) limit.
Another way to motivate the identification of the gauge theory
with the throat is to think about the absorption of massless particles
considered in the previous section. In the D-brane description,
a particle incident from the asymptotic infinity 
is converted into an excitation of the stack of D-branes, i.e. into
an excitation of the gauge theory on the world volume.
In the supergravity description, a particle incident from the asymptotic 
(large $r$) region tunnels into the $r\ll L$ region and produces an excitation
of the throat. The fact that the two different descriptions of
the absorption process give identical cross-sections supports the 
identification of excitations of $AdS_5\times S^5$ with the excited
states of the ${\cal N}=4$ SYM theory.

Another strong piece of support for this identification comes from
symmetry considerations \cite{jthroat}. The isometry group of
$AdS_5$ is $SO(2,4)$, and this is also the conformal group in
$3+1$ dimensions. In addition we have the isometries of $S^5$ which
form $SU(4)\sim SO(6)$. This group is identical to the R-symmetry of
the ${\cal N}=4$ SYM theory. After including the fermionic generators
required by supersymmetry, the full isometry supergroup of the
$AdS_5\times S^5$ background is $SU(2,2|4)$, which is identical to
the ${\cal N}=4$ superconformal symmetry.
We will see that in theories with reduced supersymmetry
the compact $S^5$ factor becomes replaced by other
compact spaces $X_5$,
but $AdS_5$ is the ``universal'' factor present in the dual
description of any large $N$ CFT and realizing the $SO(2,4)$ conformal
symmetry. One may think of these backgrounds as type IIB theory
compactified on $X_5$ down to 5 dimensions. Such Kaluza-Klein
compactifications of type IIB supergravity were extensively studied in the
mid-eighties \cite{GRW,Romans,Duff}, and special attention was devoted to
the $AdS_5\times S^5$ solution because it is
a maximally supersymmetric background \cite{SH,Kim}.
It is remarkable that these early works on compactification
of type IIB theory
were actually solving large $N$ gauge theories without knowing it.

As Maldacena has emphasized, however, it is important to go beyond
the supergravity limit and think of the $AdS_5\times X_5$ space
as a background of string theory \cite{jthroat}. Indeed, type IIB strings
are dual to the electric flux lines in the gauge theory, and
this provides a natural set-up for calculating correlation 
functions of the Wilson loops. Furthermore, if $N$ is sent to infinity
while $g_{\rm YM}^2 N$ is held fixed and finite, then there are finite
string scale corrections to the supergravity limit \cite{jthroat,US,EW}
which proceed in powers of
\be {\alpha'\over L^2} \sim \left (g_{\rm YM}^2 N \right)^{-1/2}
\ .
\ee
If we wish to study finite $N$, then there are also string loop
corrections in powers of
\be {\kappa^2\over L^8} \sim N^{-2}
\ .
\ee
As expected, taking $N$ to infinity enables us to take
the classical limit of the string theory on $AdS_5\times X_5$.
However, in order to understand the large $N$
gauge theory with finite `t Hooft coupling,
we should think of the $AdS_5\times X_5$ as the target space of a 
2-dimensional sigma model describing the classical string physics 
\cite{US}. The fact that after the compactification on $X_5$ 
the string theory is 5-dimensional supports Polyakov's idea \cite{Sasha}.
In $AdS_5$ the fifth dimension is related to
the radial coordinate and, after a change of variables
$z= L e^{-\varphi/L}$, the sigma model action turns into a special
case of the general ansatz proposed in \cite{Sasha},
\be
I = {1\over 2}\int d^2 \sigma [(\partial_i \varphi)^2 + a^2 (\varphi)
(\partial_i X^\mu)^2 + \ldots ]
\ ,
\ee
where $a(\varphi) = e^{\varphi/L}$.
It is clear, however, that the string sigma models dual to the gauge
theories are of rather peculiar nature. The new feature revealed
by the D-brane approach, which is also a major stumbling block,
is the presence of the Ramond-Ramond background fields. Little is known
to date about such 2-dimensional field theories and, in spite of
recent new insights \cite{MT}, an explicit solution is not yet available.

\subsection{Correlation functions and the bulk/boundary correspondence}

Maldacena's work provided a crucial insight that the $AdS_5\times S^5$ 
throat is the part of the 3-brane geometry that is most directly related
to the ${\cal N}=4$ SYM theory. It is important to go further, however,
and explain precisely in 
what sense the two should be identified and how physical information
can be extracted from this duality. Major strides towards
answering these questions were made in two subsequent papers 
\cite{US,EW} where
essentially identical methods for calculating correlation functions
of various operators in the gauge theory were proposed.
As we mentioned in section 2.2, even prior to \cite{jthroat} some
information about the field/operator correspondence
and about the two-point functions had been extracted from
the absorption cross-sections. The reasoning of \cite{US}
was a natural extension of these ideas. 

One may motivate the general method
as follows. When a wave is absorbed, it tunnels from the asymptotic
infinity into the throat region, and then continues to propagate
toward smaller $r$. Let us separate the 3-brane geometry into two
regions: $r\gsim L$ and $r\lsim L$. For $r\lsim L$ the metric is
approximately that of $AdS_5\times S^5$, while for $r\gsim L$
it becomes very different and eventually approaches the flat metric.
Signals coming in from large $r$ may be thought of as disturbing the
``boundary'' of $AdS_5$ at $r\sim L$, and then propagating into the
bulk. This suggests that, if we discard the $r\gsim L$ part of the
3-brane metric, then we have to cut off the radial
coordinate of $AdS_5$ at $r\sim L$, 
and the gauge theory correlation functions are related to
the response of the string theory to boundary conditions.
Guided by this idea, \cite{US} proposed
to identify the generating functional of connected
correlation functions in the gauge theory with the extremum of the
classical string action subject to the boundary conditions 
that $\phi(x^\lambda, z) = \phi_b (x^\lambda)$ at
$z=L$ (at $z=\infty$ all fluctuations are required to vanish):\foot{
As usual, in calculating the
correlation functions in a CFT
it is convenient to carry out the Euclidean continuation. On the string
theory side we then have to use the Euclidean version of $AdS_5$.} 
\be
   W[\phi_b (x^\lambda)] = S_{\phi_b (x^\lambda)} 
     \ . 
\ee 
 $W$ generates the connected Green's functions of the gauge theory
operator that corresponds to the field $\phi$ in the sense explained
in section 2.2, 
while $S_{\phi_b (x^\lambda)} $ 
is the extremum of the classical string action
subject to the boundary conditions. 
An essentially identical prescription
was also proposed in \cite{EW} with a somewhat different motivation.
If we are interested in the
correlation functions at infinite `t Hooft coupling, then the
problem of extremizing the
classical string action reduces to solving the equations of
motion in type IIB supergravity whose form is known explicitly
\cite{SH}.
Note that from the point of view of the metric (\ref{adsmetric})
the boundary conditions are imposed not at $z=0$ 
(which would be a true boundary 
of $AdS_5$) but at some finite value $z= z_{cutoff}$.
It does not matter which value it is since it can be changed by an
overall rescaling of the coordinates $(z, x^{\lambda})$ which leaves
the metric unchanged.
The physical meaning of this cut-off is that it acts as a UV
cut-off in the gauge theory \cite{US,sw}. Indeed, the radial coordinate of
$AdS_5$ is to be thought of as the effective energy scale of the gauge
theory \cite{jthroat}, and decreasing $z$ corresponds to increasing
energy. In some calculations one may remove the cut-off from the beginning
and specify the boundary conditions at $z=0$, but in others the cut-off
is needed at intermediate stages and may be removed 
only at the end \cite{freed}.

There is a growing literature on explicit calculations of correlation
functions following the proposal of \cite{US,EW}.
In these notes we will limit ourselves to a brief discussion of the
2-point functions. Their calculations show that 
the dimensions of gauge
invariant operators are determined by
the masses of the corresponding fields in $AdS_5$ \cite{US,EW}.
For scalar operators this relation is
\be
\label{dimen}
\Delta= 2+\sqrt{ 4 + (m L)^2 }\ .
\ee
Therefore, the operators in the ${\cal N}=4$ large $N$ SYM theory
naturally break up into two classes: those that correspond to the
Kaluza-Klein states of supergravity and those that correspond to 
massive string states. Since the radius of the $S^5$
is $L$, the masses of the Kaluza-Klein states are
proportional to $1/L$. Thus,
the dimensions of the corresponding operators are independent of $L$ and
therefore independent of $g_{\rm YM}^2 N$.
On the gauge theory side this
is explained by the fact that the supersymmetry protects the dimensions
of certain operators from being renormalized: they are completely determined
by the representation under the superconformal symmetry 
\cite{hw,Ferrara}.
All families of the Kaluza-Klein states, which correspond to such
BPS protected operators, were classified long ago \cite{Kim}.

On the other hand, the masses of string excitations are
$m^2 = {4 n\over \alpha'}$ where $n$ is an integer.
For the corresponding operators the formula (\ref{dimen})
predicts that the dimensions do depend on the `t Hooft coupling and,
in fact, blow up for large $g_{\rm YM}^2 N$ as
$2\left (n g_{\rm YM} \sqrt {2 N} \right )^{1/2}$.
This is a highly non-trivial prediction of the AdS/CFT duality which
has not yet been verified on the gauge theory side.

It is often stated that the gauge theory lives on the boundary of
$AdS_5$. A more precise statement is that the gauge theory corresponds
to the entire $AdS_5$, with the effective energy scale measured by
the radial coordinate. In this correspondence the bare (UV) quantities
in the gauge theory are indeed specified at the boundary of $AdS_5$.
In calculating the correlation functions it is crucial that the boundary
values of various fields in $AdS_5$ act as the sources in the
gauge theory action which couple to gauge invariant operators as in
(\ref{sint}). A similar connection arises in the calculation
of Wilson loop expectation values \cite{Malda}. 
A Wilson loop is specified by a
contour in $x^\lambda$ space placed at $z=z_{cutoff}$. 
One then looks for a minimal area surface in $AdS_5$ bounded by this 
contour and evaluates the Nambu action $I_0$ which is proportional
to the area. The semiclassical value of the Wilson loop is
then $e^{-I_0}$. This prescription, which is motivated by the duality
between fundamental strings and electric flux lines, gives interesting
results which are consistent with the conformal invariance \cite{Malda}. 
For example,
the quark-antiquark potential scales as $\sqrt{g_{\rm YM}^2 N}/|\vec x|$.
Note that this strong coupling result is different
from the weak coupling limit where we have
$V\sim g_{\rm YM}^2 N/|\vec x|$.

\subsection{Conformal field theories and Einstein manifolds}

As we mentioned above, the duality between strings on $AdS_5\times S^5$
and the ${\cal N}=4$ SYM is naturally generalized to dualities between
strings on $AdS_5\times X_5$ and other conformal gauge theories.
The 5-dimensional compact space $X_5$ is required to be a
postively curved Einstein manifold, i.e. one for which
$R_{\mu\nu}= \Lambda g_{\mu\nu}$ with $\Lambda>0$.
The number of supersymmetries in the dual gauge theory is determined
by the number of Killing spinors on $X_5$.

The simplest examples of $X_5$ are the orbifolds $S^5/\Gamma$ where 
$\Gamma$ is a discrete subgroup of $SO(6)$ \cite{ks,lnv}. In these cases
$X_5$ has the local geometry of a 5-sphere.
The dual gauge theory is the IR
limit of the world volume theory on a stack of $N$ D3-branes placed at
the orbifold singularity of $R^6/\Gamma$. Such theories
typically involve product gauge groups $SU(N)^k$ coupled to matter
in bifundamental representations \cite{dm}.

Constructions of the dual gauge theories for Einstein manifolds
$X_5$ which are not locally equivalent to $S^5$ are also possible. 
The simplest
example is the Romans compactification on
$X_5= T^{1,1}= (SU(2)\times SU(2))/U(1)$ \cite{Romans,KW}. It turns out that
the dual gauge theory is the conformal limit of 
the world volume theory on a stack of $N$ D3-branes placed at
the singularity of a certain Calabi-Yau
manifold known as the conifold. This turns out to be the ${\cal N}=1$
superconformal field theory with gauge group $SU(N)\times SU(N)$
coupled to two chiral superfields in the $({\bf N}, \overline{\bf N})$
representation
and two chiral superfields in the $(\overline{\bf N}, {\bf N})$
representation \cite{KW}.
This theory has an exactly marginal quartic superpotential 
which produces a critical line related to the radius of 
$AdS_5\times T^{1,1}$.

\section{Towards Non-conformal Gauge Theories in Four Dimensions}

In the preceding sections I hope to have convinced the reader
that type IIB strings on $AdS_5\times X_5$ shed 
genuinely new light on four-dimensional
conformal gauge theories. While many insights have already been
achieved, I am convinced that a great deal remains to be learned in
this domain.
We should not forget, however, that the prize question is whether this duality
can be extended to QCD or at least to other gauge theories which
exhibit the asymptotic freedom and confinement.
It is immediately clear that this will not be easy because, as we 
remarked in section 3, a string approach to weakly coupled gauge
theory has not yet been fully developed (the well-understood supergravity
limit describes gauge theory with very large `t Hooft coupling).
On the other hand, the asymptotic freedom makes the coupling
approach zero in the UV region \cite{GWP}.
Nevertheless, there may be some at least qualitative approaches to
non-conformal gauge theories that shed light on the essential physical
phenomena.

One such approach, proposed by Witten \cite{newWit}, builds on the
observation that thermal gauge theories are described by near-extremal
$p$-brane solutions \cite{gkp,ENT}. It is also known that the high
temperature limit of a supersymmetric gauge theory in $p+1$ dimensions
is described by non-supersymmetric gauge theory in $p$ dimensions.
Thus, 3-dimensional non-supersymmetric gauge theory
is dual to the throat region of the near-extremal 3-brane solution
which turns out to have the geometry of a black hole in 
$AdS_5$ \cite{newWit} (similar black holes were studied long ago
by Hawking and Page \cite{HP}).
Similarly, 4-dimensional non-supersymmetric gauge theory
is dual to the near-horizon
region of the near-extremal 4-brane solution \cite{newWit}.
Witten calculated the Wilson loop expectation values in these geometries
and showed that they satisfy the area law. Furthermore, calculations
of the glueball masses produce discrete spectra with strong resemblance
to the lattice simulations \cite{oog}. Unfortunately, this supergravity model
has some undesirable features as well: for example, the presence
in the geometry 
of a large $8-p$ dimensional sphere introduces 
into the spectrum families of light ``Kaluza-Klein glueballs''
which are certainly absent from the lattice results.
Presumably, the root of the problems is that the bare `t Hooft coupling
is taken to be large, while in order to achieve the conventional
continuum limit it has to be sent to zero along a renormalization
group trajectory. 

A pessimistic conclusion would be that little more can be
done at present because the supergravity approximation is supposed
to be poor at weak `t Hooft coupling.
Nevertheless, I feel that one should not give up attempts to
understand the asymptotic freedom on the string side of the duality.
In fact, some progress in this direction was recently achieved in
\cite{KTnew,JM,KTthree} following Polyakov's 
suggestion \cite{AP} on how to break supersymmetry.
Polyakov argued that a string dual of non-supersymmetric gauge theory
should have world sheet supersymmetry without space-time
supersymmetry. Examples of such theories include the type $0$
strings, which are NSR strings with a non-chiral GSO 
projection which breaks the space-time supersymmetry \cite{DH}.

There are two type $0$ theories, A and B, and both
have no space-time fermions in their spectra
but produce modular invariant
partition functions \cite{DH}.
The massless bosonic fields are  as in the corresponding
type II theory (A or B),
but with the  doubled set of the Ramond-Ramond (R-R)  fields.
The type 0 theory also contains a tachyon,
which is why it has not received much attention thus far.
In \cite{AP,KTnew} it was suggested, however, that the presence of the
tachyon does not spoil its application to large $N$ gauge theories.
A well-established route towards gauge theory is via the
D-branes which were first considered in the type 0 context in 
\cite{berg}.
Large $N$ gauge theories, which are constructed on $N$ coincident D-branes
of type 0 theory, may be shown to contain no open string
tachyons \cite{KTnew,AP}.

In \cite{KTnew} the presence of a bulk tachyon was turned into
an advantage because it gives rise to the renormalization group flow.
There the $3+1$ dimensional $SU(N)$ theory coupled to 6 adjoint massless
scalars was constructed as the low-energy description of $N$
coincident electric D3-branes of type $0$B theory.\footnote{
In the type $0$B theory the 5-form R-R field strength $F_5$
is not constrained
to be selfdual. Therefore, it is possible to have electrically or 
magnetically charged 3-branes. This should be contrasted with the situation
in the type IIB theory where the 5-form strength is constrained
to be selfdual and, thus, only the selfdual 3-branes are allowed.}
 The conjectured dual type 0
background thus carries $N$ units of electric 5-form flux.
The dilaton decouples from the $(F_5)^2$ terms in
the effective action, and the only source for it originates
from the tachyon mass term,
\be
\nabla^2 \Phi =\textstyle{ 1\over 8}   m^2
e^{{1\ov 2}\Phi} T^2
\ , \ \ \ \ \ \ \ \  m^2 = - { 2 \ov \a'} \ .
\ee
Thus, the tachyon background induces a radial variation of $\Phi$.
Since the radial coordinate is related to the energy scale of
the gauge theory, the effective coupling decreases
toward the ultraviolet. In \cite{JM,KTthree} the UV limit of the 
type $0$B background dual to the gauge theory was studied in more detail
and a solution was found where the geometry is $AdS_5\times S^5$
while the `t Hooft coupling flows logarithmically.
A calculation of the quark-antiquark potential showed qualitative
agreement with what is expected in an asymptotically free theory.

These results raise the hope that the AdS/CFT duality can indeed
be generalized to asymptotically free gauge theories.
While we are still far from constructing reliable string duals
of such theories, the availability of new ideas on this old and
difficult problem makes me hopeful that more surprises lie ahead.

\section*{Acknowledgements}
I am grateful to S. Gubser, A. Peet, A. Polyakov, A. Tseytlin and
E. Witten, my collaborators on parts of the material reviewed in these
notes. I also thank B. Kursunoglu and other organizers
of Orbis Scientiae '98, especially L. Dolan (the convener of the
string session), for sponsoring a very interesting
conference. This work  was supported in part by the NSF grant PHY-9802484 
and by the James S. McDonnell Foundation Grant No. 91-48.


\end{document}